\documentclass[prd,aps,showpacs,epsf,floats,onecolumn,10pt]{revtex4}
\usepackage{amssymb}
\usepackage{amsfonts}
\usepackage{amsmath}

\input{tcilatex}

\begin{document}

\title{Information Geometry, Inference Methods and Chaotic Energy Levels
Statistics }
\author{Carlo Cafaro}
\email{carlocafaro2000@yahoo.it}
\affiliation{Department of Physics, State University of New York at Albany-SUNY,1400
Washington Avenue, Albany, NY 12222, USA}

\begin{abstract}
In this Letter, we propose a novel information-geometric characterization of
chaotic (integrable) energy level statistics of a quantum antiferromagnetic
Ising spin chain in a tilted (transverse) external magnetic field. Finally,
we conjecture our results might find some potential physical applications in
quantum energy level statistics.
\end{abstract}

\maketitle

%\section{Introduction}

\section{Introduction}

Research on complexity \cite{gell-mann} has created a new set of ideas on
how very simple systems may give rise to very complex behaviors. Moreover,
in many cases, the "laws of complexity" have been found to hold universally,
caring not at all for the details of the system's constituents. Chaotic
behavior is a particular case of complex behavior and it will be the object
of the present work.

In this Letter we make use of the so-called Entropic Dynamics (ED) \cite%
{caticha1}. ED is a theoretical framework that arises from the combination
of inductive inference (Maximum Entropy Methods (ME), \cite{caticha2}) and
Information Geometry (IG) \cite{amari}. The most intriguing question being
pursued in ED stems from the possibility of deriving dynamics from purely
entropic arguments. This is clearly valuable in circumstances where
microscopic dynamics may be too far removed from the phenomena of interest,
such as in complex biological or ecological systems, or where it may just be
unknown or perhaps even nonexistent, as in economics. It has already been
shown that entropic arguments do account for a substantial part of the
formalism of quantum mechanics, a theory that is presumably fundamental \cite%
{caticha3}. Perhaps the fundamental theories of physics are not so
fundamental; they may just be consistent, objective ways to manipulate
information. Following this line of thought, we extend the applicability of
ED to temporally-complex (chaotic) dynamical systems on curved statistical
manifolds and identify relevant measures of chaoticity of such an
information geometrodynamical approach to chaos (IGAC).

The layout of this Letter is as follows. In section II, we present an
introduction to the main features of our IGAC. In section III, using our
information geometrodynamical approach to chaos and following the results
provided by Prosen \cite{prosen}, we propose a novel information-geometric
characterization of chaotic (integrable) energy level statistics of a
quantum antiferromagnetic Ising chain in a tilted (transverse) external
magnetic field. We emphasize that we have omitted technical details that
will appear elsewhere. However, some applications of our IGAC to low
dimensional chaotic systems can be found in our previous articles \cite%
{cafaro1, cafaro2, cafaro3, cafaro4}. Finally, in section IV we present our
final remarks.

\section{Theoretical structure of the IGAC}

IGAC\ arises as a theoretical framework to study chaos in informational
geodesic flows describing, physical, biological or chemical \ systems. It is
the information-geometric analogue of conventional geometrodynamical
approaches \cite{casetti, di bari} where the classical configuration space%
\textbf{\ }$\Gamma _{E}$\textbf{\ }is being replaced by a statistical
manifold\textbf{\ }$\mathcal{M}_{S}$\textbf{\ }with the additional
possibility of considering chaotic dynamics arising from non conformally
flat metrics (the Jacobi metric is always conformally flat, instead). It is
an information-geometric extension of the Jacobi geometrodynamics (the
geometrization of a Hamiltonian system by transforming it to a geodesic flow 
\cite{jacobi}). The reformulation of dynamics in terms of a geodesic problem
allows the application of a wide range of well-known geometrical techniques
in the investigation of the solution space and properties of the equation of
motion. The power of the Jacobi reformulation is that all of the dynamical
information is collected into a single geometric object in which all the
available manifest symmetries are retained- the manifold on which geodesic
flow is induced. For example, integrability of the system is connected with
existence of Killing vectors and tensors on this manifold. The sensitive
dependence of trajectories on initial conditions, which is a key ingredient
of chaos, can be investigated from the equation of geodesic deviation. In
the Riemannian \cite{casetti} and Finslerian \cite{di bari} (a Finsler
metric is obtained from a Riemannian metric by relaxing the requirement that
the metric be quadratic on each tangent space) geometrodynamical approach to
chaos in classical Hamiltonian systems, an active field of research concerns
the possibility\textbf{\ }of finding a rigorous relation among the sectional
curvature, the Lyapunov exponents, and the Kolmogorov-Sinai dynamical
entropy (i. e. the sum of positive Lyapunov exponents) \cite{kawabe}. Using
information-geometric methods, we have investigated in some detail the above
still open research problem \cite{cafaro2}. One of the goals of this Letter
is that of representing an additional step forward in that research
direction.

\subsection{General Formalism of the IGAC}

The IGAC is an application of ED to complex systems of arbitrary nature. ED
is a form of information-constrained dynamics built on curved statistical
manifolds $\mathcal{M}_{S}$ where elements of the manifold are probability
distributions $\left\{ P\left( X|\Theta \right) \right\} $ that are in a
one-to-one relation with a suitable set of macroscopic statistical variables 
$\left\{ \Theta \right\} $ that provide a convenient parametrization of
points on $\mathcal{M}_{S}$. The set $\left\{ \Theta \right\} $ is called
the \textit{parameter space }$\mathcal{D}_{\Theta }$ of the system.

In what follows, we schematically outline the main points underlying the
construction of an arbitrary form of entropic dynamics. First, the
microstates of the system under investigation must be defined. For the sake
of simplicity, we assume the system is characterized by an\textbf{\ }$l$%
-dimensional microspace with microstates $\left\{ X\right\} $. The main goal
of an ED model is that of inferring "macroscopic predictions" in the absence
of detailed knowledge of the microscopic nature of arbitrary complex
systems. More explicitly, with "macroscopic prediction" we mean the
knowledge of the probability function statistical parameters (expectation
values, variances, etc.). Once the microstates have been defined, we have to
select the relevant information about such microstates. In other words, we
have to select the \textit{macrospace} of the system. It is worthwhile
mentioning that the coexistence of macroscopic dynamics with microscopic
dynamics for a given physical (biological, chemical) system from a dynamical
and statistical point of view has always been a very important object of
investigation \cite{kaneto}. For example, in certain fluid systems showing
Rayleigh-Benard convection \cite{lorenz}, the macroscopic chaotic behavior
(macroscopic chaos) is a manifestation of the underlying molecular
interaction of a very large collection of molecules (microscopic or
molecular chaos). Moreover, macroscopic chaos arising from an underlying
microscopic chaotic molecular behavior has been observed in chemical
reactions. In order to study underlying dynamics in rate equations of
chemical reactions, a mesoscopic description has been adopted, which is
given by a set of transition probabilities among chemicals. In such a
description, the underlying dynamics of macroscopic motion is that one of
stochastic processes and the evolution of probability distribution of each
chemical is investigated \cite{fox}.

For the sake of the argument, we assume that the degrees of freedom\textbf{\ 
}$\left\{ x_{k}\right\} $\textbf{\ }of the microstates\textbf{\ }$\left\{
X\right\} $\textbf{\ }are Gaussian-distributed. They are defined by $2l$%
-information constraints, for example their expectation values $\mu _{k}$
and variances $\sigma _{k}$.%
\begin{equation}
\left\langle x_{k}\right\rangle \equiv \mu _{k}\text{ and }\left(
\left\langle \left( x_{k}-\left\langle x_{k}\right\rangle \right)
^{2}\right\rangle \right) ^{\frac{1}{2}}\equiv \sigma _{k}\text{.}
\end{equation}%
In addition to information constraints, each\textbf{\ }$1$-dimensional
Gaussian distribution\textbf{\ }$p_{k}\left( x_{k}|\mu _{k}\text{, }\sigma
_{k}\right) $\textbf{\ }of each degree of freedom $x_{k}$\ must satisfy the
usual normalization conditions,%
\begin{equation}
\dint\limits_{-\infty }^{+\infty }dx_{k}p_{k}\left( x_{k}|\mu _{k}\text{, }%
\sigma _{k}\right) =1\text{ }
\end{equation}%
where 
\begin{equation}
p_{k}\left( x_{k}|\mu _{k}\text{, }\sigma _{k}\right) =\left( 2\pi \sigma
_{k}^{2}\right) ^{-\frac{1}{2}}\exp \left( -\frac{\left( x_{k}-\mu
_{k}\right) ^{2}}{2\sigma _{k}^{2}}\right) \text{.}  \label{1d-gaussian}
\end{equation}%
(More correctly, we should label the Gaussian probability distribution in (%
\ref{1d-gaussian}) as\textbf{\ }$P\left( x_{k}|\mu _{k}\text{, }\sigma
_{k}\right) $, where\textbf{\ }$P$\textbf{\ }indicates a specific Gaussian
distribution selected from a generic parametric family of distributions).
Once the microstates have been defined and the relevant (linear or
nonlinear) information constraints selected, we are left with a set of $l$%
-dimensional vector probability distributions $p\left( X|\Theta \right) =%
\underset{k=1}{\overset{l}{\dprod }}$ $p_{k}\left( x_{k}|\mu _{k}\text{, }%
\sigma _{k}\right) $ encoding the relevant available information about the
system where $X$ is the $l$-dimensional microscopic vector with components $%
\left( x_{1}\text{,...,}x_{l}\right) $ and $\Theta $ is the $2l$-dimensional
macroscopic vector with coordinates $\left( \mu _{1}\text{,..., }\mu _{l}%
\text{; }\sigma _{1}\text{,..., }\sigma _{l}\right) $. The set $\left\{
\Theta \right\} $ defines the $2l$-dimensional space of macrostates of the
system, the statistical manifold $\mathcal{M}_{S}$. A measure of
distinguishability among macrostates is obtained by assigning a probability
distribution $P\left( X|\Theta \right) \ni \mathcal{M}_{S}$ to each
macrostate $\Theta $ . Assignment of a probability distribution to each
state endows $\mathcal{M}_{S}$ with a metric structure. Specifically, the
Fisher-Rao information metric $g_{\mu \nu }\left( \Theta \right) $ \cite%
{amari}, 
\begin{equation}
g_{\mu \nu }\left( \Theta \right) =\int dXp\left( X|\Theta \right) \partial
_{\mu }\log p\left( X|\Theta \right) \partial _{\nu }\log p\left( X|\Theta
\right) \text{,}  \label{fisher-rao}
\end{equation}%
with $\mu $, $\nu =1$,..., $2l$ and $\partial _{\mu }=\frac{\partial }{%
\partial \Theta ^{\mu }}$ , defines a measure of distinguishability among
macrostates on $\mathcal{M}_{S}$. The statistical manifold $\mathcal{M}_{S}$,%
\begin{equation}
\mathcal{M}_{S}=\left\{ p\left( X|\Theta \right) =\underset{k=1}{\overset{l}{%
\dprod }}p_{k}\left( x_{k}|\mu _{k}\text{, }\sigma _{k}\right) \right\} 
\text{,}
\end{equation}%
is defined as the set of probabilities $\left\{ p\left( X|\Theta \right)
\right\} $ described above where $X\in 
%TCIMACRO{\U{211d} }%
%BeginExpansion
\mathbb{R}
%EndExpansion
^{3N}$, $\Theta \in \mathcal{D}_{\Theta }=\left[ \mathcal{I}_{\mu }\times 
\mathcal{I}_{\sigma }\right] ^{3N}$. The parameter space $\mathcal{D}%
_{\Theta }$ (homeomorphic to $\mathcal{M}_{S}$) is the direct product of the
parameter subspaces $\mathcal{I}_{\mu }$ and $\mathcal{I}_{\sigma }$, where
(unless specified otherwise) $\mathcal{I}_{\mu }=\left( -\infty \text{, }%
+\infty \right) _{\mu }$ and $\mathcal{I}_{\sigma }=\left( 0\text{, }+\infty
\right) _{\sigma }$. It is worthwhile pointing out that the possible chaotic
behavior of the set of macrostates\textbf{\ }$\left\{ \Theta \right\} $%
\textbf{\ }is strictly related to the selected relevant information about
the set of microstates\textbf{\ }$\left\{ X\right\} $\textbf{\ }of the
system. In other words, the assumed Gaussian characterization of the degrees
of freedom\textbf{\ }$\left\{ x_{k}\right\} $\ of each microstate of the
system has deep consequences on the macroscopic behavior of the system
itself.

Once $\mathcal{M}_{S}$ and $\mathcal{D}_{\Theta }$ are defined, the ED
formalism provides the tools to explore dynamics driven \ on $\mathcal{M}%
_{S} $\ by entropic arguments. Specifically, given a known initial
macrostate $\Theta ^{\left( \text{initial}\right) }$ (probability
distribution), and that the system evolves to a final known macrostate $%
\Theta ^{\left( \text{final}\right) }$, the possible trajectories of the
system are examined in the ED approach using ME methods.

We emphasize ED can be derived from a standard principle of least action (of
Jacobi type). The geodesic equations for the macrovariables of the Gaussian
ED model are given by\textit{\ nonlinear} second order coupled ordinary
differential equations,%
\begin{equation}
\frac{d^{2}\Theta ^{\mu }}{d\tau ^{2}}+\Gamma _{\nu \rho }^{\mu }\frac{%
d\Theta ^{\nu }}{d\tau }\frac{d\Theta ^{\rho }}{d\tau }=0\text{.}
\label{geodesic equations}
\end{equation}%
The geodesic equations in (\ref{geodesic equations}) describe a \textit{%
reversible} dynamics whose solution is the trajectory between an initial $%
\Theta ^{\left( \text{initial}\right) }$ and a final macrostate $\Theta
^{\left( \text{final}\right) }$. The trajectory can be equally well
traversed in both directions. Given the Fisher-Rao information metric, we
can apply standard methods of Riemannian differential geometry to study the
information-geometric structure of the manifold $\mathcal{M}_{S}$ underlying
the entropic dynamics. Connection coefficients $\Gamma _{\mu \nu }^{\rho }$,
Ricci tensor $R_{\mu \nu }$, Riemannian curvature tensor $R_{\mu \nu \rho
\sigma }$, sectional curvatures $\mathcal{K}_{\mathcal{M}_{S}}$, scalar
curvature $\mathcal{R}_{\mathcal{M}_{S}}$, Weyl anisotropy tensor $W_{\mu
\nu \rho \sigma }$, Killing fields $\xi ^{\mu }$ and Jacobi fields $J^{\mu }$
can be calculated in the usual way \cite{casetti, di bari}.

To characterize the chaotic behavior of complex entropic dynamical systems,
we are mainly concerned with the signs of the scalar and sectional
curvatures of $\mathcal{M}_{S}$, the asymptotic behavior of Jacobi fields $%
J^{\mu }$ on $\mathcal{M}_{S}$, the existence of Killing vectors $\xi ^{\mu
} $ (or existence of a non-vanishing Weyl anisotropy tensor, the anisotropy
of the manifold underlying system dynamics plays a significant role in the
mechanism of instability) and the asymptotic behavior of the
information-geometrodynamical entropy (IGE) $\mathcal{S}_{\mathcal{M}_{S}}$
(see (\ref{IGE})). It is crucial to observe that true chaos is identified by
the occurrence of two features \cite{di bari}: 1) strong dependence on
initial conditions and exponential divergence of the Jacobi vector field
intensity, i.e., \textit{stretching} of dynamical trajectories; 2)
compactness of the configuration space manifold, i.e., \textit{folding} of
dynamical trajectories. The negativity of the Ricci scalar $\mathcal{R}_{%
\mathcal{M}_{S}}$,%
\begin{equation}
\mathcal{R}_{\mathcal{M}_{S}}=R_{\mu \nu \rho \sigma }g^{\mu \rho }g^{\nu
\sigma }=\sum_{\rho \neq \sigma }\mathcal{K}_{\mathcal{M}_{S}}\left( e_{\rho
}\text{, }e_{\sigma }\right) \text{,}
\end{equation}%
implies the existence of expanding directions in the configuration space
manifold $\mathcal{M}_{s}$. Indeed, since $\mathcal{R}_{\mathcal{M}_{S}}$ is
the sum of all sectional curvatures of planes spanned by pairs of
orthonormal basis elements $\left\{ e_{\rho }=\partial _{\Theta _{\rho
}}\right\} $, the negativity of the Ricci scalar is only a \textit{sufficient%
} (not necessary) condition for local instability of geodesic flow. For this
reason, the negativity of the scalar provides a \textit{strong }criterion of
local instability. Scenarios may arise where negative sectional curvatures
are present, but the positive ones could prevail in the sum so that the
Ricci scalar is non-negative despite the instability in the flow in those
directions. Consequently, the signs of $\mathcal{K}_{\mathcal{M}_{S}}$ are
of primary significance for the proper characterization of chaos.

A powerful mathematical tool to investigate the stability or instability of
a geodesic flow is the Jacobi-Levi-Civita equation (JLC equation) for
geodesic spread \cite{casetti},%
\begin{equation}
\frac{D^{2}J^{\mu }}{D\tau ^{2}}+R_{\nu \rho \sigma }^{\mu }\frac{\partial
\Theta ^{\nu }}{\partial \tau }J^{\rho }\frac{\partial \Theta ^{\sigma }}{%
\partial \tau }=0\text{.}
\end{equation}%
The JLC-equation covariantly describes how nearby geodesics locally scatter
and relates the stability or instability of a geodesic flow with curvature
properties of the ambient manifold. Finally, the asymptotic regime of
diffusive evolution describing the possible exponential increase of average
volume elements on $\mathcal{M}_{s}$ provides another useful indicator of
dynamical chaoticity. The exponential instability characteristic of chaos
forces the system to rapidly explore large areas (volumes) of the
statistical manifold. It is interesting to note that this asymptotic
behavior appears also in the conventional description of quantum chaos where
the entropy \ (von Neumann) increases linearly at a rate determined by the
Lyapunov exponents. The linear increase of entropy as a quantum chaos
criterion was introduced by Zurek and Paz \cite{zurek1}. In my
information-geometric approach a relevant quantity that can be useful to
study the degree of instability characterizing ED models is the information
geometrodynamical entropy (IGE) defined as \cite{cafaro2},%
\begin{equation}
\mathcal{S}_{\mathcal{M}_{s}}\left( \tau \right) \overset{\text{def}}{=}%
\underset{\tau \rightarrow \infty }{\lim }\log \mathcal{V}_{\mathcal{M}_{s}}%
\text{ with }\mathcal{V}_{\mathcal{M}_{s}}\left( \tau \right) =\frac{1}{\tau 
}\dint\limits_{0}^{\tau }d\tau ^{\prime }\left( \underset{\mathcal{M}_{s}}{%
\int }\sqrt{g}d^{2l}\Theta \right)  \label{IGE}
\end{equation}%
and $g=\left\vert \det \left( g_{\mu \nu }\right) \right\vert $. IGE is the
asymptotic limit of the natural logarithm of the statistical weight defined
on $\mathcal{M}_{s}$ and represents a measure of temporal complexity of
chaotic dynamical systems whose dynamics is underlined by a curved
statistical manifold. In conventional approaches to chaos, the notion of
entropy is introduced, in both classical and quantum physics, as the missing
information about the systems fine-grained state \cite{caves}. For a
classical system, suppose that the phase space is partitioned into very
fine-grained cells of uniform volume $\Delta v$, labelled by an index $j$.
If one does not know which cell the system occupies, one assigns
probabilities $p_{j}$ to the various cells; equivalently, in the limit of
infinitesimal cells, one can use a phase-space density $\rho \left(
X_{j}\right) =\frac{p_{j}}{\Delta v}$. Then, in a classical chaotic
evolution, the asymptotic expression of the information needed to
characterize a particular coarse-grained trajectory out to time $\tau $ is
given by the Shannon information entropy (measured in bits) \cite{caves},%
\begin{equation}
\mathcal{S}_{\text{classical}}^{\left( \text{chaotic}\right) }=-\int dX\rho
\left( X\right) \log _{2}\left( \rho \left( X\right) \Delta v\right)
=-\sum_{j}p_{j}\log _{2}p_{j}\sim \mathcal{K}\tau \text{.}  \label{KS}
\end{equation}%
where $\rho \left( X\right) $ is the phase-space density and $p_{j}=\frac{%
v_{j}}{\Delta v}$ is the probability for the corresponding coarse-grained
trajectory. $\mathcal{S}_{\text{classical}}^{\left( \text{chaotic}\right) }$
is the missing information about which fine-grained cell the system
occupies. The quantity $\mathcal{K}$ represents the linear rate of
information increase and it is called the Kolmogorov-Sinai entropy (or
metric entropy) ($\mathcal{K}$ is the sum of positive Lyapunov exponents, $%
\mathcal{K}=\sum_{j}\lambda _{j}$ ). $\mathcal{K}$ quantifies the degree of
classical chaos.

\section{Application of the IGAC\ to Quantum Energy Level Statistics}

The relevant indicators of chaoticity within the IGAC\ are the Ricci scalar
curvature\textbf{\ }$\mathcal{R}_{\mathcal{M}_{s}}$\textbf{\ }(or, more
correctly, the sectional curvature\textbf{\ }$\mathcal{K}_{\mathcal{M}_{S}}$%
), the Jacobi vector field intensity $J_{\mathcal{M}_{S}}$\textbf{\ }and the
IGE\textbf{\ }$\mathcal{S}_{\mathcal{M}_{s}}$ once the line element on the
curved statistical manifold\textbf{\ }$\mathcal{M}_{s}$\textbf{\ }underlying
the entropic dynamics has been specified. In what follows, selected two
special line elements, we focus exclusively on the asymptotic temporal
behavior of the IGE\textbf{\ }$\mathcal{S}_{\mathcal{M}_{s}}$\textbf{\ }%
related to the ED arising from them. 

We apply the IGAC to study the entropic dynamics on curved statistical
manifolds induced by classical probability distributions of common use in
the study of regular and chaotic quantum energy level statistics. In doing
so, we suggest an information-geometric characterization of a special class
of regular and chaotic quantum energy level statistics.

As we said previously, we have omitted technical details that will appear
elsewhere. However, our previous works (especially (\cite{cafaro2})) may be
very useful references in order to clarify the following application. Recall
that the theory of quantum chaos (quantum mechanics of systems whose
classical dynamics are chaotic) is not primarily related to few-body
physics. Indeed, in real physical systems such as many-electron atoms and
heavy nuclei, the origin of complex behavior is the very strong interaction
among many particles. To deal with such systems, a famous statistical
approach has been developed which is based upon the Random Matrix Theory
(RMT). The main idea of this approach is to neglect the detailed description
of the motion and to treat these systems statistically bearing in mind that
the interaction among particles is so complex and strong that generic
properties are expected to emerge. Once again, this is exactly the
philosophy underlining the ED approach to complex dynamics. It is known that
the asymptotic behavior of computational costs and entanglement entropies of
integrable and chaotic Ising spin chains are very different \cite{prosen} .
Here Prosen considered the question of time efficiency implementing an
up-to-date version of the t-DMRG for a family of Ising spin $\frac{1}{2}$
chains in arbitrary oriented magnetic field, which undergoes a transition
from integrable (transverse Ising) to nonintegrable chaotic regime as the
magnetic field is varied. An integrable (regular) Ising chain in a general
homogeneous transverse magnetic field is defined through the Hamiltonian $%
\mathcal{H}_{\text{regular}}\left( 0\text{, }2\right) $, where%
\begin{equation}
\mathcal{H}\left( h_{x}\text{, }h_{y}\right) =\underset{j=0}{\overset{n-2}{%
\sum }}\sigma _{j}^{x}\sigma _{j+1}^{x}+\underset{j=0}{\overset{n-1}{\sum }}%
\left( h^{x}\sigma _{j}^{x}+h^{y}\sigma _{j}^{y}\right) \text{.}
\label{ising-hamiltonian}
\end{equation}%
In this case, the computational cost shows a polynomial growth in time, $%
D_{\varepsilon }^{\left( \text{regular}\right) }\left( t\right) \overset{%
\tau \rightarrow \infty }{\propto }\tau $, while the entanglement entropy is
characterized by logarithmic growth,%
\begin{equation}
\mathcal{S}_{\text{regular}}\left( 0\text{, }2\right) =\mathcal{S}_{\text{%
von Neumann}}^{\left( 0\text{, }2\right) }\overset{\tau \rightarrow \infty }{%
\propto }c\log \tau +c^{\prime }  \label{regularentropy}
\end{equation}%
The constant\textbf{\ }$c$\ depends exclusively on the value of the fixed
transverse magnetic field intensity\textbf{\ }$B_{\perp }$, while\textbf{\ }$%
c^{\prime }$\textbf{\ }depends on\textbf{\ }$B_{\perp }$\textbf{\ }and on
the choice of the initial local operators of finite index used to calculate
the operator space entanglement entropy. Instead, a quantum chaotic Ising
chain in a general homogeneous tilted magnetic field is defined through the
Hamiltonian $\mathcal{H}_{\text{chaotic}}\left( 1\text{, }1\right) $, where $%
\mathcal{H}$ is defined in (\ref{ising-hamiltonian}). In this case, the
computational cost shows an exponential growth in time, $D_{\varepsilon
}^{\left( \text{chaotic}\right) }\left( t\right) \overset{\tau \rightarrow
\infty }{\propto }\exp \left( \mathcal{K}_{q}\tau \right) $, while the
entanglement entropy is characterized by linear growth,%
\begin{equation}
\mathcal{S}_{\text{chaotic}}\left( 1\text{, }1\right) =\mathcal{S}_{\text{%
von Neumann}}^{\left( 1\text{, }1\right) }\overset{\tau \rightarrow \infty }{%
\propto }\mathcal{K}_{q}\tau \text{.}  \label{QDE}
\end{equation}%
The quantity\textbf{\ }$\mathcal{K}_{q}$\textbf{\ }is a constant,
asymptotically independent of the number of indexes of the initial local
operators used to calculate the operator space entropy, that depends only on
the Hamiltonian evolution and not on the details of the initial state
observable or error measures, and can be interpreted as a kind of quantum
dynamical entropy.

It is well known the quantum description of chaos is characterized by a
radical change in the statistics of quantum energy levels \cite{casati}. The
transition to chaos in the classical case is associated with a drastic
change in the statistics of the nearest-neighbor spacings of quantum energy
levels. In the regular regime the distribution agrees with the Poisson
statistics while in the chaotic regime the Wigner-Dyson distribution works
very well. Uncorrelated energy levels are characteristic of quantum systems
corresponding to a classically regular motion while a level repulsion (a
suppression of small energy level spacing) is typical for systems which are
classically chaotic. A standard quantum example is provided by the study of
energy level statistics of an Hydrogen atom in a strong magnetic field. It
is known that level spacing distribution (LSD) is a standard indicator of
quantum chaos \cite{haake}. It displays characteristic level repulsion for
strongly nonintegrable quantum systems, whereas for integrable systems there
is no repulsion due to existence of conservation laws and quantum numbers.
In \cite{prosen}, the authors calculate the LSD of the spectra of $\mathcal{H%
}_{\text{regular}}\left( 0\text{, }2\right) $ and $\mathcal{H}_{\text{chaotic%
}}\left( 1\text{, }1\right) $. They find that for $\mathcal{H}_{\text{regular%
}}\left( 0\text{, }2\right) $, the nearest neighbor LSD is described by a
Poisson distribution. For $\mathcal{H}_{\text{chaotic}}\left( 1\text{, }%
1\right) $, they find the nearest neighbor LSD is described by a
Wigner-Dyson distribution. Therefore, they conclude that $\mathcal{H}_{\text{%
regular}}\left( 0\text{, }2\right) $ and $\mathcal{H}_{\text{chaotic}}\left(
1\text{, }1\right) $ indeed represent generic regular and quantum chaotic
systems, respectively. We will encode the relevant information about the
spin-chain in a suitable composite-probability distribution taking account
of the quantum spin chain and the configuration of the external magnetic
field in which they are immersed.

In the ME\ method \cite{caticha2}, the selection of relevant variables is
made on the basis of intuition guided by experiment; it is essentially a
matter of trial and error. The variables should include those that can be
controlled or experimentally observed, but there are cases where others must
also be considered. Our objective here is to choose the relevant
microvariables of the system and select the relevant information concerning
each one of them. In the integrable case, the Hamiltonian $\mathcal{H}_{%
\text{regular}}\left( 0\text{, }2\right) $ describes an antiferromagnetic
Ising chain immersed in a transverse homogeneous magnetic field $\vec{B}_{%
\text{transverse}}=B_{\perp }$ $\hat{B}_{_{\perp }}$ with the level spacing
distribution of its spectrum given by the Poisson distribution%
\begin{equation}
p_{\text{Poisson}}\left( x_{A}|\mu _{A}\right) =\frac{1}{\mu _{A}}\exp
\left( -\frac{x_{A}}{\mu _{A}}\right) \text{,}  \label{poisson-integrable}
\end{equation}%
where the microvariable $x_{A}$ is the spacing of the energy levels and the
macrovariable $\mu _{A}$ is the average spacing. The chain is immersed in
the \textit{transverse} magnetic field which has just one component $%
B_{\perp }$ in the Hamiltonian $\mathcal{H}_{\text{regular}}\left( 0\text{, }%
2\right) $. We translate this piece of information in our IGAC formalism,
coupling the probability (\ref{poisson-integrable}) to an exponential bath $%
p_{B}^{\left( \text{exponential}\right) }\left( x_{B}|\mu _{B}\right) $
given by%
\begin{equation}
p_{B}^{\left( \text{exponential}\right) }\left( x_{B}|\mu _{B}\right) =\frac{%
1}{\mu _{B}}\exp \left( -\frac{x_{B}}{\mu _{B}}\right) \text{,}
\end{equation}%
where the microvariable $x_{B}$ is the intensity of the magnetic field and
the macrovariable $\mu _{B\text{ }}$is the average intensity. More
correctly, $x_{B}$\ should be the energy arising from the interaction of the 
\textit{transverse} magnetic field with the spin $\frac{1}{2}$\ particle
magnetic moment, $x_{B}=\left\vert -\vec{\mu}\cdot \vec{B}\right\vert
=\left\vert -\mu B\cos \varphi \right\vert $\ where $\varphi $\ is the tilt
angle. For the sake of simplicity, let us set $\mu =1$, then in the
transverse case $\varphi =0$\ and therefore $x_{B}=B\equiv B_{\perp }$. This
is our best guess and we justify it by noticing that the magnetic field
intensity is indeed a relevant quantity in this experiment (see equation (%
\ref{regularentropy})) and its components (intensity) are quantities that
are varied during the transitions from integrable to chaotic regimes. In the
regular regime, we say the magnetic field intensity is set to a well-defined
value $\left\langle x_{B}\right\rangle =\mu _{B}$. Furthermore, notice that
the Exponential distribution is identified by information theory as the
maximum entropy distribution if only one piece of information (the
expectation value) is known. Finally, the chosen composite probability
distribution $P^{\left( \text{integrable}\right) }\left( x_{A}\text{, }%
x_{B}|\mu _{A}\text{, }\mu _{B}\right) $ encoding relevant information about
the system is given by,%
\begin{equation}
P^{\left( \text{integrable}\right) }\left( x_{A}\text{, }x_{B}|\mu _{A}\text{%
, }\mu _{B}\right) =\frac{1}{\mu _{A}\mu _{B}}\exp \left[ -\left( \frac{x_{A}%
}{\mu _{A}}+\frac{x_{B}}{\mu _{B}}\right) \right] \text{.}  \label{p-exp}
\end{equation}%
Again, we point out that our probability (\ref{p-exp}) is our best guess
and, of course, must be consistent with numerical simulations and
experimental data in order to have some merit. We point out that equation (%
\ref{p-exp}) is not fully justified from a theoretical point of view, a
situation that occurs due to the lack of a systematic way to select the
relevant microvariables of the system (and to choose the appropriate
information about such microvariables). Let us denote $\mathcal{M}_{S}^{%
\text{integrable}}$ the two-dimensional curved statistical manifold
underlying our information geometrodynamics. The line element $ds_{\text{%
integrable}}^{2}$ on $\mathcal{M}_{S}^{\text{integrable}}$ is given by,%
\begin{equation}
ds_{\text{integrable}}^{2}=ds_{\text{Poisson}}^{2}+ds_{\text{Exponential}%
}^{2}=\frac{1}{\mu _{A}^{2}}d\mu _{A}^{2}+\frac{1}{\mu _{B}^{2}}d\mu _{B}^{2}%
\text{.}  \label{regular case}
\end{equation}%
Applying our IGAC (see (\cite{cafaro2})) to the line element in (\ref%
{regular case}), we obtain polynomial growth in $\mathcal{V}_{\mathcal{M}%
_{s}}^{\text{integrable}}$ and logarithmic IGE growth, 
\begin{equation}
\mathcal{V}_{\mathcal{M}_{s}}^{\left( \text{integrable}\right) }\left( \tau
\right) \overset{\tau \rightarrow \infty }{\propto }\exp (c_{IG}^{\prime
})\tau ^{c_{IG}}\text{, }\mathcal{S}_{\mathcal{M}_{s}}^{\left( \text{%
integrable}\right) }\left( \tau \right) \overset{\tau \rightarrow \infty }{%
\propto }c_{IG}\log \tau +c_{IG}^{\prime }\text{.}  \label{polinomio}
\end{equation}%
The quantity\textbf{\ }$c_{IG}$\textbf{\ }is a constant proportional to the
number of Exponential probability distributions in the composite
distribution used to calculate the IGE and\textbf{\ }$c_{IG}^{\prime }$\ is
a constant that depends on the values assumed by the statistical
macrovariables\textbf{\ }$\mu _{A}$\textbf{\ }and\textbf{\ }$\mu _{B}$.
Equations in (\ref{polinomio}) may be interpreted as the
information-geometric analogue of the computational complexity $%
D_{\varepsilon }^{\left( \text{regular}\right) }\left( \tau \right) $\ and
the entanglement entropy $\mathcal{S}_{\text{regular}}\left( 0\text{, }%
2\right) $\ defined in standard quantum information theory, respectively. We
cannot state they are the same since we are not fully justifying, from a
theoretical standpoint, our choice of the composite probability (\ref{p-exp}%
).

In the chaotic case, the Hamiltonian $\mathcal{H}_{\text{chaotic}}\left( 1%
\text{, }1\right) $ describes an antiferromagnetic Ising chain immersed in a
tilted homogeneous magnetic field $\vec{B}_{\text{tilted}}=B_{\perp }$ $\hat{%
B}_{\perp }+B_{\parallel }$ $\hat{B}_{\parallel }$ with the level spacing
distribution of its spectrum given by the Poisson distribution $p_{\text{%
Wigner-Dyson}}\left( x_{A}^{\prime }|\mu _{A}^{\prime }\right) $%
\begin{equation}
p_{\text{Wigner-Dyson}}\left( x_{A}^{\prime }|\mu _{A}^{\prime }\right) =%
\frac{\pi x_{A}^{\prime }}{2\mu _{A}^{\prime 2}}\exp \left( -\frac{\pi
x_{A}^{\prime 2}}{4\mu _{A}^{\prime 2}}\right) \text{.}
\label{wigner-dyson(chaotic)}
\end{equation}%
where the microvariable $x_{A}^{\prime }$ is the spacing of the energy
levels and the macrovariable $\mu _{A}^{\prime }$is the average spacing. The
chain is immersed in the \textit{tilted} magnetic vector field which has two
components $B_{\perp }$ and $B_{\parallel }$ in the Hamiltonian $\mathcal{H}%
_{\text{chaotic}}\left( 1\text{, }1\right) $. We translate this piece of
information in our IGAC formalism, coupling the probability (\ref%
{wigner-dyson(chaotic)}) to a Gaussian $p_{B}^{\left( \text{Gaussian}\right)
}\left( x_{B}^{\prime }|\mu _{B}^{\prime }\text{, }\sigma _{B}^{\prime
}\right) $ given by,%
\begin{equation}
p_{B}^{\left( \text{Gaussian}\right) }\left( x_{B}^{\prime }|\mu
_{B}^{\prime }\text{, }\sigma _{B}^{\prime }\right) =\frac{1}{\sqrt{2\pi
\sigma _{B}^{\prime 2}}}\exp \left( -\frac{\left( x_{B}^{\prime }-\mu
_{B}^{\prime }\right) ^{2}}{2\sigma _{B}^{\prime 2}}\right) \text{.}
\end{equation}%
where the microvariable $x_{B}^{\prime }$ is the intensity of the magnetic
field, the macrovariable $\mu _{B\text{ }}^{\prime }$is the average
intensity of the magnetic energy arising from the interaction of the \textit{%
tilted} magnetic field with the spin $\frac{1}{2}$\ particle magnetic
moment, and $\sigma _{B}^{\prime }$ is its covariance:\ during the
transition from the integrable to the chaotic regime, the magnetic field
intensity is being varied (experimentally). It is being tilted and its two
components ($B_{\perp }$ and $B_{\parallel }$) are being varied as well. Our
best guess based on the experimental mechanism that drives the transitions
between the two regimes is that magnetic field intensity ( actually the
microvariable $\mu B\cos \varphi $) is Gaussian-distributed (two
macrovariables) during this change. In the chaotic regime, we say the
magnetic field intensity is set to a well-defined value $\left\langle
x_{B}^{\prime }\right\rangle =\mu _{B}^{\prime }$\ with covariance $\sigma
_{B}^{\prime }=\sqrt{\left\langle \left( x_{B}^{\prime }-\left\langle
x_{B}^{\prime }\right\rangle \right) ^{2}\right\rangle }$. Furthermore,%
\textbf{\ }the Gaussian distribution is identified by information theory as
the maximum entropy distribution if only the expectation value and the
variance are known. Therefore, the chosen composite probability distribution 
$P^{\left( \text{chaotic}\right) }\left( x_{A}^{\prime }\text{, }%
x_{B}^{\prime }|\mu _{A}^{\prime }\text{, }\mu _{B}^{\prime }\text{, }\sigma
_{B}^{\prime }\text{ }\right) $ encoding relevant information about the
system is given by,%
\begin{equation}
P^{\left( \text{chaotic}\right) }\left( x_{A}^{\prime }\text{, }%
x_{B}^{\prime }|\mu _{A}^{\prime }\text{, }\mu _{B}^{\prime }\text{, }\sigma
_{B}^{\prime }\text{ }\right) =\frac{\pi \left( 2\pi \sigma _{B}^{\prime
2}\right) ^{-\frac{1}{2}}}{2\mu _{A}^{\prime 2}}x_{A}^{\prime }\exp \left[
-\left( \frac{\pi x_{A}^{\prime 2}}{4\mu _{A}^{\prime 2}}+\frac{\left(
x_{B}^{\prime }-\mu _{B}^{\prime }\right) ^{2}}{2\sigma _{B}^{\prime 2}}%
\right) \right] \text{.}
\end{equation}%
Let us denote $\mathcal{M}_{S}^{\left( \text{chaotic}\right) }$ the
three-dimensional curved statistical manifold underlying our information
geometrodynamics. The line element $ds_{\text{chaotic}}^{2}$ on $\mathcal{M}%
_{S}^{\left( \text{chaotic}\right) }$ is given by,%
\begin{equation}
ds_{\text{chaotic}}^{2}=ds_{\text{Wigner-Dyson}}^{2}+ds_{\text{Gaussian}%
}^{2}=\frac{4}{\mu _{A}^{\prime 2}}d\mu _{A}^{\prime 2}+\frac{1}{\sigma
_{B}^{\prime 2}}d\mu _{B}^{\prime 2}+\frac{2}{\sigma _{B}^{\prime 2}}d\sigma
_{B}^{\prime 2}\text{.}  \label{chaotic case}
\end{equation}%
Applying our IGAC \ (see (\cite{cafaro2})) to the line element in (\ref%
{chaotic case}), we obtain exponential growth for $\mathcal{V}_{\mathcal{M}%
_{s}}^{\text{chaotic}}$ and linear IGE growth 
\begin{equation}
\mathcal{V}_{\mathcal{M}_{s}}^{\left( \text{chaotic}\right) }\left( \tau
\right) \overset{\tau \rightarrow \infty }{\propto }C_{IG}\exp \left( 
\mathcal{K}_{IG}\tau \right) \text{, }\mathcal{S}_{\mathcal{M}_{s}}^{\left( 
\text{chaotic}\right) }\left( \tau \right) \overset{\tau \rightarrow \infty }%
{\propto }\mathcal{K}_{IG}\tau \text{.}  \label{exponential}
\end{equation}%
The constant\textbf{\ }$C_{IG}$\textbf{\ }encodes information about the
initial conditions of the statistical macrovariables parametrizing elements
of $\mathcal{M}_{S}^{\left( \text{chaotic}\right) }$. The constant $\mathcal{%
K}_{IG}$,%
\begin{equation}
\mathcal{K}_{IG}\overset{\tau \rightarrow \infty }{\approx }\frac{d\mathcal{S%
}_{\mathcal{M}_{s}}\left( \tau \right) }{d\tau }\overset{\tau \rightarrow
\infty }{\approx }\underset{\tau \rightarrow \infty }{\lim }\left[ \frac{1}{%
\tau }\log \left( \left\Vert \frac{J_{\mathcal{M}_{S}}\left( \tau \right) }{%
J_{\mathcal{M}_{S}}\left( 0\right) }\right\Vert \right) \right] \overset{%
\text{def}}{=}\lambda _{J}\text{,}
\end{equation}%
is the model parameter of the chaotic system and depends on the temporal
evolution of the statistical macrovariables. It plays the role of the
standard Lyapunov exponent of a trajectory and it is, in principle, an
experimentally observable quantity. The quantity\textbf{\ }$J_{\mathcal{M}%
_{S}}\left( \tau \right) $\ is the Jacobi field intensity and\textbf{\ }$%
\lambda _{J}$\textbf{\ }may be considered the information-geometric analogue
of the leading Lyapunov exponent in conventional Hamiltonian systems. Given
an explicit expression of\textbf{\ }$\mathcal{K}_{IG}$\textbf{\ }in terms of
the observables $\mu _{\text{A }}^{\prime }$\textbf{\ }and\textbf{\ }$\mu _{%
\text{B }}^{\prime }$\textbf{\ }and\textbf{\ }$\sigma _{\text{B}}^{\prime }$%
, a clear understanding of the relation between the IGE (or\textbf{\ }$%
\mathcal{K}_{IG}$) and the entanglement entropy (or\textbf{\ }$\mathcal{K}%
_{q}$) becomes the key point that deserves further study.\textbf{\ }%
Equations in (\ref{exponential}) are the information-geometric analogue of
the computational complexity $D_{\varepsilon }^{\left( \text{chaotic}\right)
}\left( \tau \right) $ and the entanglement entropy $\mathcal{S}_{\text{%
chaotic}}\left( 1\text{, }1\right) $ defined in standard quantum information
theory, respectively. This result is remarkable, but deserves a deeper
analysis in order to be fully understood.

One of the major limitations of our findings is the lack of a detailed
account of the comparison of the theory with experiment. This point will be
one of our primary concerns in future works. However, some considerations
may be carried out at the present stage. The experimental observables in our
theoretical models are the statistical macrovariables characterizing the
composite probability distributions. In the integrable case, where the
coupling between a Poisson distribution and an Exponential one is considered,%
\textbf{\ }$\mu _{\text{A }}$\textbf{\ }and\textbf{\ }$\mu _{\text{B }}$%
\textbf{\ }are the experimental observables. In the chaotic case, where the
coupling between a Wigner-Dyson distribution and a Gaussian is considered,%
\textbf{\ }$\mu _{\text{A }}^{\prime }$\textbf{\ }and\textbf{\ }$\mu _{\text{%
B }}^{\prime }$\textbf{\ }and\textbf{\ }$\sigma _{\text{B}}^{\prime }$%
\textbf{\ }play the role of the experimental observables. We believe one way
to test our theory may be that of determining a numerical estimate of the
leading Lyapunov exponent\textbf{\ }$\lambda _{\text{max}}$\textbf{\ }or the
Lyapunov spectrum for the Hamiltonian systems under investigation directly
from experimental data (measurement of a time series) and compare it to our
theoretical estimate for\textbf{\ }$\lambda _{J}$\textbf{\ }\cite{wolf}.%
\textbf{\ }However, we are aware that it may be extremely hard to evaluate
Lyapunov exponents numerically. Otherwise, knowing that the mean values of
the positive Lyapunov exponents are related to the Kolmogorov-Sinai (KS)
dynamical entropy, we suggest to measure the KS entropy\textbf{\ }$\mathcal{K%
}$\textbf{\ }directly from a time signal associated to a suitable
combination of our experimental observables and compare it to our indirect
theoretical estimate for\textbf{\ }$\mathcal{K}_{IG}$\textbf{\ }from the
asymptotic behaviors of our statistical macrovariables \cite{procaccia}. We
are aware that the ground of our discussion is quite qualitative. However,
we hope that with additional study, especially in clarifying the relation
between the IGE and the entanglement entropy, our theoretical
characterization presented in this Letter will find experimental support in
the future. Therefore, the statement that our findings may be relevant to
experiments verifying the existence of chaoticity and related dynamical
properties on a macroscopic level in energy level statistics in chaotic and
regular quantum spin chains is purely a conjecture at this stage.

\section{Conclusions}

In this Letter, we proposed a theoretical information-geometric framework
suitable to characterize chaotic dynamical behavior of complex systems on
curved statistical manifolds. Specifically, an information-geometric
characterization of regular and chaotic quantum energy level statistics
appearing in a quantum Ising spin chain in external magnetic field was
presented. It is worthwhile emphasizing the following points:\ the
statements that spectral correlations of classically integrable systems are
well described by Poisson statistics and that quantum spectra of classically
chaotic systems are universally correlated according to Wigner-Dyson
statistics are conjectures, known as the BGS (Bohigas-Giannoni-Schmit, \cite%
{bohigas} and BTG (Berry-Tabor-Gutzwiller, \cite{gutzwiller}) conjectures,
respectively. These two conjectures are very important in the study of
quantum chaos, however their validity finds some exceptions. Several other
cases may be considered. For instance, chaotic systems having a spectrum
that does not obey a Wigner-Dyson distribution may be considered. A chaotic
system can also have a spectrum following a Poisson, semi-Poisson, or other
types of critical statistics \cite{garcia}. Moreover, integrable systems
having a spectrum that does not obey a Poisson distribution may be
considered as well. For instance, the Harper model would represent such a
situation. Moreover, it is worthwhile pointing out that not every chaotic
system characterized by entropy-like quantities growing linearly in time has
a spectrum described by a Wigner-Dyson distribution. Well-known examples
presenting such a situation are the cat maps \cite{gu} and the famous kicked
rotator \cite{izrailev} where its spectrum follows a Poisson distribution in
cylinder representation and a Wigner-Dyson in torus representation but the
properties of entropy-like quantities are the same in both representations
(at least classically). All these cases are not discussed in our
characterization.

Therefore, at present stage, because of the above considerations and because
of the lack of experimental\textbf{\ }evidence in support of our theoretical
construct, we can only conclude that the IGAC might find some potential
applications in certain regular and chaotic dynamical systems and this
remains only a conjecture. However, we hope that our work convincingly shows
that this information-geometric approach may be considered a serious effort
trying to provide a unifying criterion of chaos of both classical and
quantum varieties, thus deserving further research and developments.

\begin{acknowledgments}
I am grateful to Saleem Ali, Ariel Caticha and Adom Giffin for very useful
discussions and for their previous collaborations. I extend thanks to Cedric
Beny, Michael Frey and Jeroen Wouters for their interest and/or useful
comments on my research during the NIC@QS07 in Erice, Ettore Majorana
Centre. Finally, I sincerely thank the two anonymous Referees for
constructive criticism and for very helpful suggestions.
\end{acknowledgments}

\end{document}